# Microwave losses of bulk $CaC_6$

G. Cifariello[a], E. Di Gennaro[a], G. Lamura[b], A. Andreone[*a]

N. Emery[c], C. Hérold[c], J. F. Marêché[c], P. Lagrange[c]

[a]*CNISM and Department of Physics, University of Naples "Federico II", 80125 Naples, Italy*

[b]*CNR-INFM Coherentia and Department of Physics, University of Naples "Federico II", 80125 Naples, Italy*

[c]*Laboratoire de Chimie du Solide Minéral-UMR 7555, Université Henri Poincaré Nancy I, B.P. 239, 54506 Vandœuvre-lès-Nancy Cedex, France.*

**Abstract**

We report a study of the temperature dependence of the surface resistance $R_S$ in the graphite intercalated compound (GIC) $CaC_6$, where superconductivity at 11.5 K was recently discovered. Experiments are carried out using a copper dielectrically loaded cavity operating at 7 GHz in a "hot finger" configuration. Bulk $CaC_6$ samples have been synthesized from highly oriented pyrolytic graphite. Microwave data allows to extract unique information on the quasiparticle density and on the nature of pairing in superconductors. The analysis of $R_S(T)$ confirms our recent experimental findings that $CaC_6$ behaves as a weakly-coupled, fully gapped, superconductor.

## 1. Introduction

Elucidation of the nature of the superconducting state in the newly discovered $CaC_6$ compound [1, 2] with $T_C$ at 11.5 K is of importance both for fundamental understanding and for applied research. Two relevant issues are the nature of the superconducting pairing mechanism and the symmetry of the order parameter. There is a mounting experimental evidence from different spectroscopic tools, London penetration depth [3], tunnelling [4], and specific heat [5], that the superconducting order parameter does follow a conventional s-wave symmetry, and presents a ratio $2\Delta(0)/k_BT_C$ – where $\Delta(0)$ is the zero temperature energy gap – close to the BCS prediction. In the electromagnetic response, the characteristic signature of a gapped superconductor is an exponential behaviour of the microwave absorption, represented by the surface resistance $R_S$, at low temperatures. We have measured the microwave properties of bulk $CaC_6$ synthesized from highly oriented pyrolytic graphite [6]. The data presented here have been taken on a platelike c-axis oriented polycrystal having a roughly squared shape of maximum size 2.5x2.5 mm$^2$ and thickness of 0.1 mm.

## 2. Experimental

For measuring the surface resistance of the GIC sample, we used a perturbation method based on a open-ended dielectric single-crystal sapphire puck resonator operating at the resonant frequency of 7 GHz. The resonator enclosure is made of oxygen-free high conductivity copper, whereas the sample holder is a low loss sapphire rod, placed at the centre of the cavity in close proximity to the dielectric crystal. The cavity is excited with a transverse electric $TE_{011}$ mode, which induces a-b plane screening currents in the sample. The sapphire puck (8 mm height and 16 mm diameter) is separated from the copper wall by a thin sapphire spacer (6 mm height and 2.5 mm diameter). By using a micrometer screw, the position of the sample placed on the sapphire rod, and therefore the puck-to-

———
[*] Corresponding author. Tel.: +390817682547; fax: +390812391821; e-mail: andreone@unina.it.

sample distance, can be changed, in order to get the maximum sensitivity. The ability to move the sample makes the sensitivity variable, allowing to measure surface resistance values ranging between Ohms and µOhms. The apparatus allows any sample between 1 and 10 mm$^2$ to be placed in the cavity. The overall resonator is taken under vacuum using a copper can, which includes a double layer µ-metal shield, and inserted in a liquid helium cryostat. The technique works in the transmission mode, with microwave power coupled into and out of the cavity by stainless steel cryogenic coaxial lines terminating in loops. Measurements are performed with a Hewlett Packard 8720C vectorial network analyzer.

## 3. Results and Discussion

The microwave surface resistance $R_S(T)$ of bulk $CaC_6$ is shown in Fig. 1. Because $CaC_6$ is highly reactive when exposed to oxygen, the sample was accurately cleaved in a glove box with an inert atmosphere before the measurement. The measured normal state $R_S$ value is rather low, yielding a residual value of the normal state resistivity close to 5 µΩ·cm. This is consistent with d.c. values reported in literature [7]. As far as the surface resistance approaches $T_C$, it starts to rapidly drop as temperature decreases. However at low temperature $R_S$ tends to saturate, reaching a residual value more or less below 3 K, certainly due to surface effects. In spite of the good quality of the sample data cannot be consistently fitted in the overall temperature range within a BCS framework, since the level of extrinsic losses is too high. For $T$ less than $T_c/2$, in conventional superconductors the surface resistance behaviour can be phenomenologically described using the standard BCS exponential dependence, after subtracting to the data a residual term $R_{res}$ related to the extrinsic losses [8].

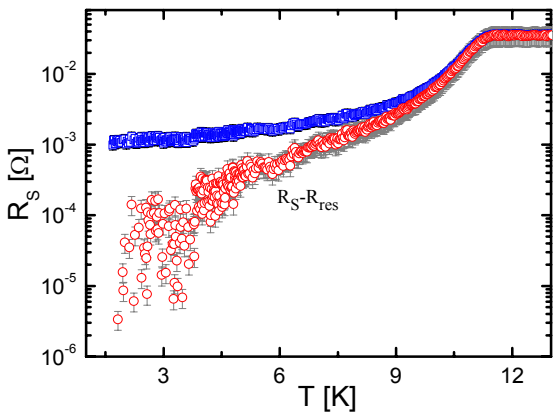

Fig. 1. $R_S$ vs $T$ for the $CaC_6$ bulk sample. The sharp superconducting transition around 11.5 K is clearly visible. Data obtained subtracting a constant term at the lowest temperatures are also shown.

This is done in fig. 2, where the quantity $R_S - R_{res}$ is displayed as a function of $T_C/T$ at low temperatures. Here $R_{res}=R_S(T_{min})$ is chosen. Once the residual value is subtracted, the low temperature data does show an exponential temperature dependence, according to $R_S \propto exp\text{-}(\Delta(0)/k_BT)$. A straight line behaviour is clearly seen at low $T$, that is an unambiguous and direct signature of the superconducting gap. Data represented in Fig. 2 represent therefore a clear evidence of a fully gapped superconductor. A fit has been performed for $T_c/T$ ranging between 2 and 4, because of the large scattering of data at the lowest temperatures, that gives a strong coupling ratio $2\Delta(0)/k_BT_C = (3.6 \pm 0.5)$. The resulting gap $\Delta(0)$ is $(1.7 \pm 0.3)$ meV, that agrees with the value obtained from penetration depth measurements [3, 9].

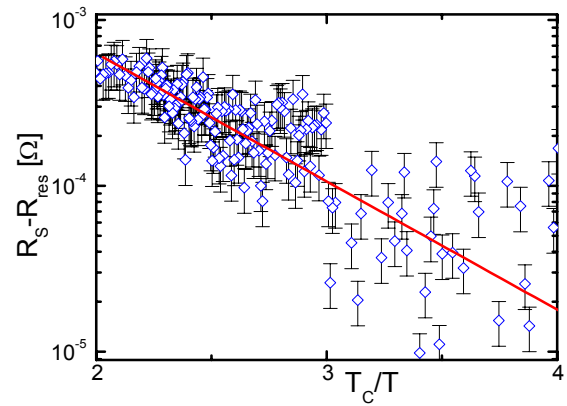

Fig. 2. $(R_S-R_{res})$ vs $T_c/T$ is plotted below $T_c/2$ on a semi-log scale. The continuous line represents the exponential behavior predicted by the BCS theory.

In summary, we have presented a study of the surface resistance in a bulk $CaC_6$ sample. In spite of the fact that its surface properties in the microwave region are far from being optimized, the observed temperature dependence, cleared of extrinsic effects, can be well explained within a conventional theory. A single-valued and finite gap throughout the entire temperature range below $T_C$ can be extracted from the data, consistent with recent measurements of the superfluid density performed on the same sample.

## References

[1] T. E. Weller *et al.*, Nature Phys. **1**, 39(2005).
[2] N. Emery *et al.*, Phys. Rev. Lett. **95**, 087003 (2005).
[3] G. Lamura *et al.*, Phys. Rev. Lett. **96**, 107008 (2006).
[4] N. Bergeal *et al.*, cond-mat/0603443 (2006).
[5] J. S. Kim *et al.*, Phys. Rev. Lett. **96**, 217002 (2006).
[6] N. Emery *et al.*, J. Solid State Chem. **178**, 2947(2005).
[7] A. Gauzzi *et al.*, cond-mat/0604208 (2006).
[8] M. Tinkham, *Introduction to Superconductivity*, McGraw-Hill, New York (1996)
[9] G. Lamura *et al.*, this conference.